\documentclass[
aip,amsmath,amssymb,reprint,floatfix]{revtex4-1}
\usepackage[T1]{fontenc}
\usepackage[latin9]{inputenc}
\usepackage{color}
\definecolor{page_backgroundcolor}{rgb}{1, 1, 1}
\pagecolor{page_backgroundcolor}
\usepackage{array}
\usepackage{enumitem}
\usepackage{amsmath}
\usepackage{graphicx}
\usepackage{esint}
\usepackage{subfig}

\draft 

\begin{document}


\title{Stochastic embedding DFT: theory and application to p-nitroaniline
in water} 



\author{Wenfei Li}%
\affiliation{Department of Chemistry and Biochemistry, University of California, Los Angeles California 90095, USA}
\author{Ming Chen}%
\affiliation{Affiliation: Department of Chemistry, University of California and Materials Science Division, Lawrence Berkeley National Laboratory, Berkeley, California 94720, USA}
\author{Eran Rabani}%
\affiliation{Affiliation: Department of Chemistry, University of California and Materials Science Division, Lawrence Berkeley National Laboratory, Berkeley, California 94720, USA}
\author{Roi Baer}%
\affiliation{Affiliation: Fritz Haber Center for Molecular Dynamics, Institute of Chemistry, The Hebrew University of Jerusalem, Jerusalem 91904, Israel}
\author{Daniel Neuhauser}%
\affiliation{Department of Chemistry and Biochemistry, University of California, Los Angeles California 90095, USA}


\date{\today}

\begin{abstract}
Over this past decade, we combined the idea of stochastic resolution
of identity with a variety of electronic structure methods. In our
stochastic Kohn-Sham DFT method, the density is an average over multiple
stochastic samples, with stochastic errors that decrease as the inverse
square root of the number of sampling orbitals. Here we develop a
stochastic embedding density functional theory method (se-DFT) that
selectively reduces the stochastic error (specifically on the forces)
for a selected sub-system(s). The motivation, similar to that of other
quantum embedding methods, is that for many systems of practical interest
the properties are often determined by only a small sub-system. In
stochastic embedding DFT two sets of orbitals are used: a deterministic
one associated with the embedded subspace, and the rest which is described
by a stochastic set. The method is exact in the limit of large number
of stochastic samples. We apply se-DFT to study a p-nitroaniline molecule
in water, where the statistical errors in the forces on the system
(the p-nitroaniline molecule) are reduced by an order of magnitude
compared with non-embedding stochastic DFT.
\end{abstract}


\maketitle 
DFT (Density Functional Theory) traditionally follows the Kohn-Sham
scheme where a set of one-particle equations is solved self-consistently.
For large systems the solution of these equations scales as ${\rm O}(N_{e}^{2})-{\rm O}(N_{e}^{3})$
with the number of electrons $N_{e},$ so there is a lot of interest
in variants that scale linearly with system size. Such methods include
orbital-free DFT with density-dependent kinetic energy functionals,
\cite{Witt2018,Lehtomaki2014}; linear-scaling approaches where the
system is split into parts that are woven together via constraints\cite{lee1998linear};
as well as embedding techniques where an inner part is treated by
DFT and an outer part by orbital-free DFT\cite{wesolowski1993frozen,wesolowski2015frozen,iannuzzi2006density}.

Previously, we developed stochastic DFT (sDFT), a method that can
be viewed as a bridge between Kohn-Sham DFT and orbital-free DFT\cite{Baer2013}.
Instead of computing Kohn-Sham orbitals for all occupied states, we
apply a Chebyshev filter to a few stochastic orbitals\cite{Baer2013},
and extract the density from these filtered orbitals, circumventing
the time-consuming diagonalization step. This approach is exact in
the limit of infinitely many stochastic samples and gives useful results
even for a small number of samplings. 

In a follow-up work \cite{neuhauser2014communication,arnon2017equilibrium,chen2019overlapped}
we have shown how to reduce the standard deviation in sDFT (and therefore
accelerate the convergence), using a method we label stochastic-fragment
DFT (sf-DFT). Here, instead of sampling stochastically the full density,
we sample stochastically only the difference between the full density
and a zero-order density which is easy to calculate. The difference
is generally small, thereby reducing the fluctuations.

Here, we develop an alternate method whereby a given sub-region is
embedded. Essentially, this sub-region is treated deterministically
while the rest of the system is treated stochastically (this is a
simplifying view and the more precise methodology is described later).
The motivation for this method is that, for many realistic systems,
only a subsystem is of particular importance. The idea of embedding
was widely adopted to treat such systems. In most embedding methods,
the sub-system of interest is calculated at higher level theories,
while the rest is treated with less accurate but more efficient methods
to reduce computational cost\cite{doi:10.1021/ct300544e,Tamukong2014,hegely2016exact,culpitt2017communication,gomes2008calculation,loco2017hybrid,wesolowski2015frozen,sun2016quantum,wesolowski1993frozen,wesolowski2013non,goodpaster2010exact,goodpaster2011embedded,fux2010accurate,schnieders2018accurate}.
Our stochastic density functional theory embedding method adopts an
analogous strategy, except that here the larger stochastic region
embeds the smaller deterministic part.

An attractive feature of the stochastic embedding method is that the
errors due to the embedding are numerically controlled, since in the
limit of infinitely many stochastic samplings the method is exact.
As such, there is no residual arbitrariness due to the choice of an
embedding potential.

This paper is organized as follows. Sec. I presents the theory, and
the practical algorithm is reviewed in Sec. II. In Sec. III a practical
system is studied, embedding of a dye (p-nitroaniline) in 216 water
molecules. Discussion and possible extensions follow in Sec. IV.

\section{Theory}

\subsection{Stochastic DFT}

We first review stochastic DFT as developed in our previous works\cite{Baer2013,chen2019overlapped,doi:10.1021/ct300544e,neuhauser2014communication}.

In DFT, the key component is the electron density $\rho(\boldsymbol{r})$,
which we express as the trace of a Heaviside step function: 
\begin{center}
\begin{equation}
\frac{\rho(\boldsymbol{r})}{2}=\langle\boldsymbol{r}|\Theta(\mu-H)|\boldsymbol{r}\rangle
\end{equation}
\par\end{center}

\noindent \begin{flushleft}
where we assume spin-unpolarized DFT. Here, $\mu$ is the electron
chemical potential, determined by ensuring the correct total number
of electrons: 
\par\end{flushleft}

\begin{center}
\begin{equation}
N_{e}=\int\rho(\boldsymbol{r})dr,\label{eq:Ne}
\end{equation}
\par\end{center}

\noindent and the one-body Hamiltonian is $H=-\frac{1}{2}\nabla^{2}+v(\boldsymbol{r}),$
where we introduced the is the effective one-electron potential due
to the the nuclear ($v_{N})$ electron-electron Coulomb interaction
$(v_{H})$ and exchange-correlation $(v_{XC})$ parts. We assume for
simplicity that the exchange-correlation (and therefore the total
effective) potential depends on the local density, $v=v\left[\rho\right]$. 

In the usual deterministic formulations of Kohn-Sham DFT, the electron
density is expressed as the sum over one-electron states, and the
total number of electrons is determined by the occupation number of
each state. Thus, the Heaviside filter becomes a projection to the
occupied subspace: 
\begin{center}
\begin{equation}
\Theta(\mu-H)=\sum_{i\le N_{{\rm occ}}}|\psi_{i}\rangle\langle\psi_{i}|.
\end{equation}
\par\end{center}

\noindent where we introduced the number of occupied orbitals ($N_{{\rm occ}}=N_{e}/2$).
The one-electron orbitals $\psi_{i}$ are obtained by diagonalization
of the effective one-electron Hamiltonian matrix, resulting in a nominal
$N_{e}^{3}$ scaling of Kohn-Sham DFT. Expectation values of one-electron
operators are obtained from the occupied states: 
\begin{center}
\begin{equation}
\langle A\rangle=\sum_{i\le N_{{\rm occ}}}\langle\psi_{i}|A|\psi_{i}\rangle.
\end{equation}
\par\end{center}

In stochastic Kohn-Sham DFT, on the other hand, we use a set of stochastic
orbitals, $\xi(r)$, with the property that: 
\begin{center}
\begin{equation}
\left\{ |\xi\rangle\langle\xi|\right\} _{\xi}=I,
\end{equation}
\par\end{center}

\noindent where the curly brackets stand for averaging over all stochastic
orbitals. Inserting the identity operator in the expression for $\rho(\boldsymbol{r})$,
the electron density is thus expressible as: 
\begin{center}
\begin{equation}
\rho(\boldsymbol{r})=\left\{ \langle\boldsymbol{r}|\Theta^{\frac{1}{2}}|\xi\rangle\langle\xi|\Theta^{\frac{1}{2}}|\boldsymbol{r}\rangle\right\} _{\xi}=\left\{ |\xi_{\mu}(\boldsymbol{r})|^{2}\right\} _{\xi},
\end{equation}
\par\end{center}

\noindent where we abbreviate $\Theta^{\frac{1}{2}}\equiv\Theta^{\frac{1}{2}}(\mu-H)$,
and $\xi_{\mu}\equiv\Theta^{\frac{1}{2}}\xi.$ 

The filtered stochastic orbitals are linear combinations of all occupied
states with random coefficients: 
\begin{center}
\begin{equation}
\xi_{\mu}(\boldsymbol{r})=\sum_{i\le N_{{\rm occ}}}c_{i}\phi_{i}(\boldsymbol{r})\label{eq:ximu}
\end{equation}
\par\end{center}

\noindent where $c_{i}=\langle\phi_{i}|\xi\rangle,$ so
\begin{center}
\begin{equation}
\left\{ c_{i}^{*}c_{j}\right\} =\delta_{ij}.
\end{equation}
\par\end{center}

Similarly, the expectation values of any one-particle operator $D$
is: 
\begin{center}
\begin{equation}
\langle D\rangle=\left\{ \langle\xi_{\mu}|D|\xi_{\mu}\rangle\right\} _{\xi}\approx\frac{1}{N_{s}}\sum_{\xi}\langle\xi_{\mu}|D|\xi_{\mu}\rangle.
\end{equation}
\par\end{center}

\noindent Here $N_{s}$ is the number of stochastic orbitals used
in practice. As in any stochastic method, the expectation value, obtained
as the average of a finite number of samples, will have an associated
stochastic error which is proportional to $1/\sqrt{N_{s}}$. The actual
number of stochastic orbitals is chosen based on the required level
of precision.

In practice the method relies on the fact that the calculation of
each stochastic vector scales only linearly with system size. Specifically,
we use a smooth Heaviside function, $\Theta(\mu-H)=\frac{1}{2}{\rm erfc}\left(\beta\left(\mu-H\right)\right)$
where $\beta$ needs to be much larger than the inverse band gap.
The smooth theta function is then expressed as a finite sum of Chebyshev
polynomials, $\Theta^{\frac{1}{2}}=\sum_{n}a_{n}(\mu)T_{n}(H_{{\rm scaled}}),$
where $H_{{\rm scaled}}$ is a scaled Hamiltonian with eigenvalues
in the range $\left[-1,1\right]$ and $a_{n}(\mu)$ are the Chebyshev
coefficients of $\left(\frac{1}{2}{\rm erfc}\left(\beta\left(\mu-H\right)\right)\right)^{\frac{1}{2}}$.
Therefore 
\begin{equation}
|\xi_{\mu}\rangle=\sum_{n}a_{n}(\mu)|\xi^{n}\rangle,
\end{equation}
where the Chebyshev vectors are obtained recursively, $|\xi^{n}\rangle=2H_{{\rm scaled}}|\xi^{n-1}\rangle-|\xi^{n-2}\rangle,$
and $|\xi^{n=0}\rangle\equiv|\xi\rangle$.

The Chebyshev expansion makes it possible to analytically determine
the chemical potential. Specifically, expand $\Theta=\sum_{n}b_{n}(\mu)T_{n}(H_{{\rm scaled}})$,
where $b_{n}(\mu)$ are the Chebyshev coefficients of $\frac{1}{2}{\rm erfc}\left(\beta\left(\mu-H\right)\right)$.
Then, using Eq. (\ref{eq:Ne}) gives:
\begin{equation}
\frac{N_{e}}{2}=\frac{1}{N_{s}}\sum_{\xi}\langle\xi_{\mu}|\Theta(\mu-H)|\xi_{\mu}\rangle=\sum_{n}b_{n}(\mu)R_{n}\label{eq:Neres}
\end{equation}
where $R_{n}=N_{s}^{-1}\sum_{\xi}\langle\xi|T_{n}\left(H_{{\rm scaled}}\right)|\xi\rangle.$
Therefore, $\mu$ is varied until Eq. (\ref{eq:Neres}) is fulfilled.

Next we turn to embedding, first traditional and then stochastic.

\subsection{DFT Embedding}

As in other embedding methods, the motivation for stochastic DFT embedding
is that in many practical applications the properties of a system
depend on much smaller sub-system(s), such as defects in semiconductors
or active sites of proteins. Often, we do not even care for the rest
of the system except the embedded part. Even when a quantum-mechanical
treatment of the rest of the system (the environment) is necessary,
the level of precision required is usually not as high as that of
the sub-system(s) of interest. 

A key component in all types of embedding methods is the specific
quantity through which the properties of the environment is conveyed
to the sub-system(s). In DFT embedding, the quantity is the electron
density of the entire system \cite{sun2016quantum}. Specifically,
subset A denotes the sub-system of interest, and subset B is associated
with the environment (the precise meaning of these two subsets would
be flexible). The total electron density is therefore partitioned
$\rho=\rho_{A}+\rho_{B}$ . 

In traditional deterministic DFT embedding one then derives an approximate
functional for the two regions that captures the energy of the environment
as well as its interaction with the sub-system(s) of interest. Solving
this equation for orbitals in subset A is equivalent to solving an
ordinary Kohn-Sham equation with an extra external potential due to
the embedding. There are different choices based on a requirement
that the orbitals of the sub-systems(s) should be orthogonal to the
orbitals of the environment. In practice, this term can be approximated\cite{sun2016quantum,wesolowski1993frozen},
or the orthogonality can be incorporated explicitly\cite{doi:10.1021/ct300544e,Tamukong2014,khait2012orthogonality,hegely2016exact,culpitt2017communication,tamukong2016accurate}. 

In our stochastic embedding DFT, there is the same loose overall goal
as in deterministic embedding, i.e., the different treatment of a
smaller system and the environment. However, the methods are quite
different. In our approach, the embedded space is treated determinstically
and the other (environment) is treated stochastically but otherwise
the treatment is exact. Therefore, the only sense in which embedding
is approximate here is numerical, i.e., if we use enough stochastic
orbitals the results are exact. The two spaces (system and environment)
see the same overall Hamiltonian, and there is no uncontrolled ansatz.

Specifcially, using the same language of partitioning space to parts,
we separate the total electron density into two parts, abbreviating
\begin{center}
\begin{align}
\frac{\rho(\boldsymbol{r})}{2} & =\langle\boldsymbol{r}|\Theta^{\frac{1}{2}}\Theta^{\frac{1}{2}}|\boldsymbol{r}\rangle\nonumber \\
 & =\langle\boldsymbol{r}|\Theta^{\frac{1}{2}}P\Theta^{\frac{1}{2}}|\boldsymbol{r}\rangle+\langle\boldsymbol{r}|\Theta^{\frac{1}{2}}Q\Theta^{\frac{1}{2}}||\boldsymbol{r}\rangle\nonumber \\
 & \equiv\frac{\rho_{A}(\boldsymbol{r})}{2}+\frac{\rho_{B}(\boldsymbol{r})}{2}.
\end{align}
\par\end{center}

The first term in the splitting projects onto the ``A'' subspace,
defined by its basis
\begin{center}
\begin{equation}
\hat{P}=\sum_{i\in A}|\chi_{i}\rangle\langle\chi_{i}|,\,\,\,\,\,\,\,\langle\chi_{i}|\chi_{j}\rangle=\delta_{ij}.
\end{equation}
\par\end{center}

\noindent Therefore, 
\begin{center}
\begin{equation}
\frac{1}{2}\rho_{A}(\boldsymbol{r})=\sum_{i\in A}\langle\boldsymbol{r}|\Theta^{\frac{1}{2}}|\chi_{i}\rangle\langle\chi_{i}|\Theta^{\frac{1}{2}}|\boldsymbol{r}\rangle=\sum_{i\in A}|\chi_{i,\mu}(\boldsymbol{r})|^{2},\label{eq:rho_A}
\end{equation}
\par\end{center}

\noindent where $\chi_{i,\mu}=\Theta^{\frac{1}{2}}(\mu-H)\chi_{i}$.
Note that the $\chi_{i}(\boldsymbol{r})$ basis does not have to be
related directly to the molecular orbitals but would typically be
a made from a set of atomic orbitals in a given region, although there
is a lot of freedom in the definition. For example, in the example
studied later we choose a set of local Gaussian atomic orbitals on
each atom in the embedded subsystem (labeled as $\phi_{i}(\boldsymbol{r})$,
$i\in A$) and then orthogonalize them, to produce $\chi_{i}(\boldsymbol{r})=\sum_{j}(S^{-\frac{1}{2}})_{ij}\phi_{j}(\boldsymbol{r})$
where $S_{ij}=\langle\phi_{i}|\phi_{j}\rangle$ is the overlap matrix
of the embedded-part atomic orbitals.

The second term in the splitting is associated with $Q\equiv I-P$,
the orthogonal projection to the other (``B'') subspace. Since $Q^{2}=Q$
and inserting the identity operator $I=\left\{ |\xi\rangle\langle\xi|\right\} _{\xi}$
we get
\begin{center}
\begin{align}
\frac{1}{2}\rho_{B}(\boldsymbol{r})&=\langle\boldsymbol{r}|\Theta^{\frac{1}{2}}QQ\Theta^{\frac{1}{2}}|\boldsymbol{r}\rangle \nonumber \\
&=\left\{ \langle\boldsymbol{r}|\Theta^{\frac{1}{2}}Q|\xi\rangle\langle\xi|Q\Theta^{\frac{1}{2}}|\boldsymbol{r}\rangle\right\} _{\xi} \nonumber \\
&=\left\{ \Big[|\bar{\xi}_{\mu}(\boldsymbol{r})|^{2}\right\} _{\xi}
\end{align}
\par\end{center}

\noindent where
\begin{center}
\begin{equation}
|\bar{\xi}_{\mu}\rangle\equiv\Theta^{\frac{1}{2}}\left(\mu-H\right)Q|\xi\rangle\label{eq:xi_mu}
\end{equation}
\par\end{center}

\noindent is obtained by two consecutive projections: first a random
orbital is projected to the space orthogonal to the embedded $P$
part, and the result is then projected to the occupied space of the
full system.

Thus we reach the main embedding expression, the separation of the
density into two parts, 

\begin{equation}
\frac{1}{2}\rho(\boldsymbol{r})=\sum_{i\in A}|\chi_{i,\mu}(\boldsymbol{r})|^{2}+\frac{1}{N_{s}}\sum_{\xi}|\bar{\xi}_{\mu}(\boldsymbol{r})|^{2},\label{eq:rho_mu_embedded}
\end{equation}
one associated with the deterministic subspace and one with an orthogonal
stochastic part. The two parts are connected through the application
of the density matrix operator, $\Theta\left(\mu-H\right)$, since
the potential in the Hamiltonian depends on the density, $v=v[\rho]$,
and the density is a mixture of stochastic and deterministic parts.

An important feature of the algorithm is that the deterministic and
stochastic orbitals, $\chi_{i\mu}$ and $\bar{\xi}_{\mu}$, that make
up the density (Eq. (\ref{eq:rho_mu_embedded})), are not orthogonal
\textendash{} neither among themselves nor to each other. The orthogonality
of the $P$ and $Q$ spaces reflects in the orthogonality of the original
$\chi_{i}$ and $Q\xi$ functions, but that orthogonality is lost
when we act on $\chi_{i}$ and on $Q\xi$ with $\Theta^{\frac{1}{2}}(\mu-H)$
in Eqs. (\ref{eq:rho_A}),(\ref{eq:xi_mu}). Further note that, as
mentioned, the same overall Hamiltonian (and therefore the same $\Theta^{\frac{1}{2}}(\mu-H)$
) is used in preparing both the stochastic and deterministic orbitals,
i.e., they are treated on equal footing.

\section{Algorithm}

The overall stochastic embedding DFT method is then quite similar
to the stochastic DFT algorithm: 
\begin{enumerate}
\item Generate $N_{s}$ stochastic orbitals: $\xi(\boldsymbol{r})=\pm1/\sqrt{d^{3}r}$
where $d^{3}r$ is the volume element associated with the grid. Also
create a reasonable initial density $\rho(\boldsymbol{r})$ which
integrates to the correct number of valence electrons. 
\item Determine the one-particle effective potential and Hamiltonian $H=T+v[\rho]$ 
\item For each stochastic orbital, project out the components along the
atomic basis functions, i.e., prepare $\bar{\xi}=Q\xi$: 
\end{enumerate}
\begin{center}
$\bar{\xi}(\boldsymbol{r})=\xi(\boldsymbol{r})-\sum_{i}c_{i}\chi_{i}(\boldsymbol{k})$ 
\par\end{center}

where 
\begin{center}
$c_{i}=\int\chi_{i}\boldsymbol{(r})\xi(\boldsymbol{r})d\boldsymbol{r.}$ 
\par\end{center}
\begin{enumerate}[resume]
\item Determine the correct chemical potential $\mu$ as the one that integrates
correctly the total density, i.e., from Eq. (\ref{eq:Neres}) where
now
\begin{equation}
R_{n}=\sum_{i\in A}\langle\chi_{i}|T_{n}\left(H_{{\rm scaled}}\right)|\chi_{i}\rangle+\frac{1}{N_{s}}\sum_{\xi}\langle\bar{\xi}|T_{n}\left(H_{{\rm scaled}}\right)|\bar{\xi}\rangle,
\end{equation}

i.e., the residues and therefore the constraint on the integrated
density include both the deterministic and stochastic parts.
\item Chebyshev filter the orthogonalized atomic basis functions as well
as the projected stochastic functions: 
\end{enumerate}
\begin{center}
$|\chi_{i,\mu}\rangle=\Theta^{\frac{1}{2}}(\mu-H)|\chi_{i}\rangle$ 
\par\end{center}

\begin{center}
$|\bar{\xi}_{\mu}\rangle=\Theta^{\frac{1}{2}}(\mu-H)|\bar{\xi\rangle},$ 
\par\end{center}
\begin{enumerate}[resume]
\item Calculate the charge densities for this $\mu$ from Eq. (\ref{eq:rho_mu_embedded}).
\item Reiterate steps $2-6$ until the density does not vary, i.e., SCF
convergence is reached. 
\end{enumerate}
With the filtered atomic basis functions and filtered stochastic orbitals
, the expectation value of any one-particle operator $D$ is: 
\begin{center}
\begin{equation}
\langle D\rangle=\sum_{i}\langle\chi_{i,\mu}|D|\chi_{i,\mu}\rangle+\frac{1}{N_{s}}\langle\bar{\xi}_{\mu}|D|\bar{\xi}_{\mu}\rangle.
\end{equation}
\par\end{center}

The algorithm is therefore very similar to the original stochastic
DFT approach. The only differences are that (i) in addition to the
stochastic functions one also needs to project the density matrix
(or more precisely $\Theta^{\frac{1}{2}}\left(\mu-H\right)$) on the
deterministic basis making the embedded part, and (ii) for the stochastic
part, we now project out the deterministic part (i.e., apply $Q$)
before filtering with the Chebyshev expansion of $\Theta^{\frac{1}{2}}\left(\mu-H\right)$
. 

\section{Computational details}

We applied the stochastic density functional embedding method to study
a realistic case of embedding, i.e., a dye in water. The dye was a
p-nitroaniline molecule, and it was embedded in 216 water molecules. 

\subsection{Structure preparation }

The configuration of the system was obtained from snapshots of molecular
dynamic simulations with Gromacs 5\cite{berendsen1995gromacs}. The
dynamics simulations used a generalized amber force field with charges
from AM1-BCC for p-nitroaniline\cite{wang2004development} and TIP4P
with allowed flexibility of bond/bend for water\cite{jorgensen1985temperature}.

\begin{figure}
\begin{centering}
\includegraphics[width=8cm]{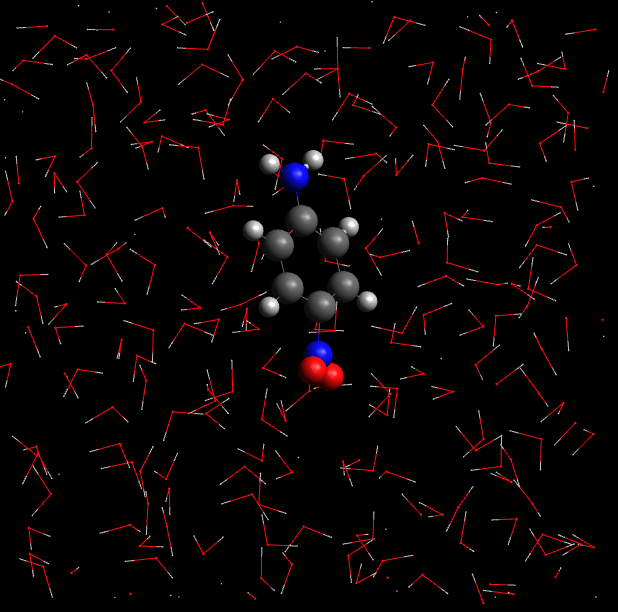} 
\par\end{centering}
\caption{\label{fig:configuration}Configuration of the p-nitroaniline/water
system used in DFT calculation. The p-nitroaniline molecule is represented
by ball and sticks, and water molecules are represented by wire-frames.}
\end{figure}

The simulation steps for the preparation of the configuration were
standard, involving first a high temperature equilibrations of the
p-nitroaniline, followed by NVT simulations at room temperature, and
then NPT equilibration at room temperature and pressure. We then ran
the MD sampled a specific configurations after several nsec. The configuration
was used for subsequent DFT calculations and is shown in Figure \ref{fig:configuration}.

\subsection{Stochastic embedding DFT details}

For the DFT calculation, we imposed periodic boundary conditions.
A plane wave expansion was used, with atomic norm-conserving Troullier-Martins
pseudopotentials replacing the core-valence interaction. An LDA functional
was used. An $88^{3}$ grid with a spacing of 0.402 atomic units was
used, while the plane wave kinetic-energy cutoff was 15 Hartree. The
inverse-temperature-like parameter $\beta$ in the Heaviside function
${\rm erfc}\left(\beta\left(\mu-H\right)\right)$ was set at $\beta=$0.03
${\rm Hartree^{-1}},$ requiring 1173 Chebyshev propagations in acting
with $\Theta^{\frac{1}{2}}(\mu-H)$.

For the embedding basis sets we used a Gaussian double-zeta basis
optimized for pseudopotentials, as given in the Quickstep\cite{vandevondele2005quickstep}
data set. \footnote{We specifically used the DZVP-GTH-PADE basis for all atoms, in the
CONFINED variant where available. The basis-set is available from
https://github.com/SINGROUP/pycp2k /blob/master/examples/BASIS\_SET }

Three sets of calculations were carried out. The first used $N_{s}=96$
stochastic orbitals, without embedding. The second used the same $N_{s}=96$
stochastic functions, but supplemented them with the double zeta atomic
basis set for all 16 atoms belonging to the p-nitroaniline molecule.
This deterministic basis set for the dye contained 160 functions (an
average of 10 functions per atom). The atomic functions were orthogonalized,
giving rise to 160 orthogonal $\chi_{i}(\boldsymbol{r})$ functions.
Therefore a total of 256 functions was employed (160 deterministic
and 96 stochastic).

Since the number of deterministic functions in the second, embedded,
set of calculations was quite large, we also compared the second set
with a third set where all functions were stochastic, and where 256
functions were used, i.e., the same overall number as in the second
set. Thus, the numerical effort, mostly associated with the Chebyshev
application of $\Theta^{\frac{1}{2}},$ is similar in the second and
third sets.

Each set of calculations was repeated ten times, with different random
seeds for generating the stochastic orbitals, to obtain the standard
deviation of the stochastic approaches.

To benchmark our results, we also performed a conventional deterministic
Kohn-Sham DFT calculation which matches the results of Quantum-Espresso.\cite{giannozzi2017advanced} 

\section{Results }

The most time-consuming step in the stochastic DFT formulation is
the application of the Chebyshev filter. Therefore, as mentioned,
the time required to perform calculation with embedding and $N_{s}=96$
(second set) is comparable to that with no embedding and $N_{s}=256$
(third set), and is about $2.5$ times the time required to perform
calculation with no embedding and $N_{s}=96$ (first set). Indeed,
in practice about 14 core hours per SCF iteration (on a cluster with
2.5 GHZ nodes) were needed for the second and third sets, with about
6 core hours for the first set. The deterministic set required about
11 core hours per iterations. In all cases, 30 DIIS iterations were
used for full SCF convergence. 

Next, we compared the total energy per electron obtained from the
three sets of calculations with that obtained from the deterministic
calculation, as well as the individual contribution from Hartree and
exchange-correlation energies. The comparison is shown in Table \ref{tab:dft_energies}.
The results show good overall agreement between the stochastic and
deterministic calculations. Most importantly the standard deviations
of energies is not affected by embedding, because the dominant contribution
comes from the 216 water molecules. The standard deviation decreases
of course by increasing $N_{S}$.

\begin{table}
\begin{centering}
\begin{tabular}{|>{\centering}p{1.6cm}|>{\centering}p{1.6cm}|>{\centering}p{1.6cm}|>{\centering}p{1.6cm}|>{\centering}p{1.3cm}|}
\hline 
& $N_{s}=96$ without embedding & $N_{s}=96$ with embedding & $N_{s}=256$ without embedding & Determi-nistic\tabularnewline
\hline 
\hline 
Total energy per electron & -2.115 $\pm$ 0.0015 & -2.115 $\pm$ 0.0016 & -2.117 $\pm$ 0.0008 & -2.117\tabularnewline
\hline 
Hartree energy per electron & 1.098 $\pm$ 0.0023 & 1.098 $\pm$ 0.0025 & 1.100 $\pm$ 0.0016 & 1.103\tabularnewline
\hline 
Exchange-correlation energy per electron & -0.521 $\pm$ 0.0004 & -0.521 $\pm$ 0.0004 & -0.521 $\pm$ 0.0002 & -0.522\tabularnewline
\hline 
\end{tabular}
\par\end{centering}
\caption{\label{tab:dft_energies}DFT energies per electron, in Hartree. For
the stochastic calculations, the energies were obtained as average
of ten calculations, and the standard deviation is included. As clearly
seen, embedding does not affect the accuracy of the overall energies.}
\end{table}

\begin{figure}
\begin{centering}
\subfloat[$N_{s}=96$, without embedding]{\begin{centering}
\includegraphics[width=6cm]{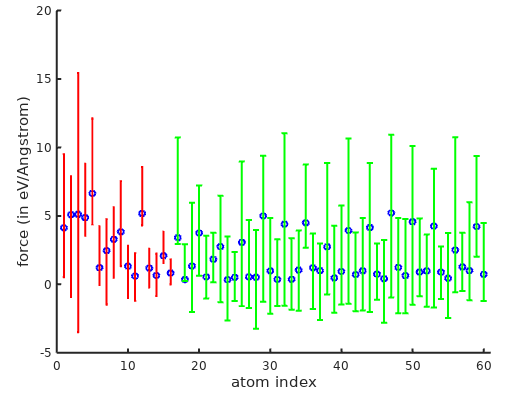} 
\par\end{centering}
}
\par\end{centering}
\begin{centering}
\subfloat[$N_{s}=96$, with embedding]{\begin{centering}
\includegraphics[width=6cm]{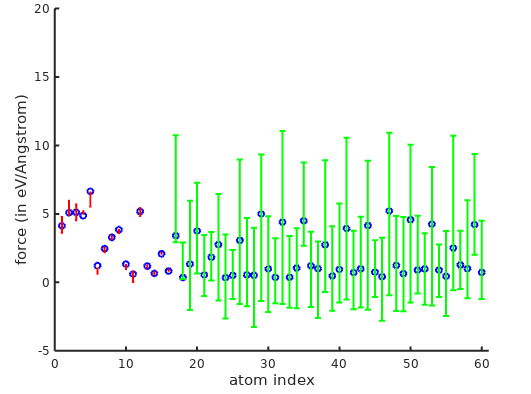} 
\par\end{centering}
}
\par\end{centering}
\begin{centering}
\subfloat[$N_{s}=256$, without embedding]{\begin{centering}
\includegraphics[width=6cm]{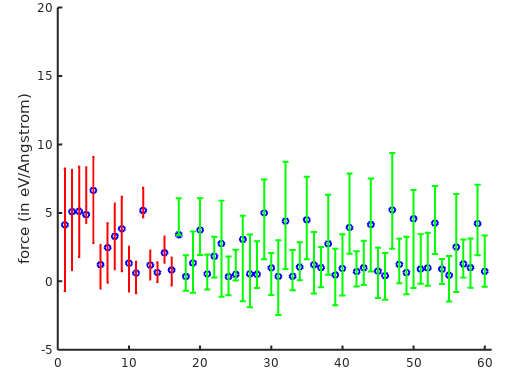} 
\par\end{centering}
}
\par\end{centering}
\caption{\label{fig:forces_fig}Forces and their standard deviations for 60
(out of the 664 overall) atoms, in ${\rm eV/\mathring{A}}$. The first
16 atoms in the plot are the p-nitroaniline molecule, and the other
44 are from water molecules. In the second set, the atoms of the p-nitroaniline
are embedded, and in the first and third set they are not. The blue
circles denote the forces calculated by deterministic DFT, while the
center of each bar refers to the average stochastic force. Red error
bars are associated with the atoms of the p-nitroaniline molecule,
and green bars are used for the standard deviation of the forces on
the water atoms.}
\end{figure}

The effect of embedding comes to play in quantities relating to the
embedded region, and the main such quantity is the force on each atom.
Figure \ref{fig:forces_fig} shows the overall forces and their standard
deviations for 60 out of the 644 atoms in the sample; the first 16
are the embedded dye, and the other 44 are from the water molecules.
The figure clearly shows that, with embedding, the forces on the embedded
atoms have much higher accuracy (much smaller standard deviation)
than the forces on the water molecule.

As a side remark note that the forces on the non-embedded atoms have
the same overall statistical fluctuations as without embedding. We
know from previous studies that for large systems like the present
one the stochastic errors are independent of size, and depend only
on the number of stochastic samples. Thus, the approach presented
here would not deteriorate with the overall size of the full system.

Coming back to the embedded system (the dye), the higher accuracy
on the dye forces is shown more quantitatively in Table \ref{tab:forces_table},
where the standard deviations of the forces on the dye is an order
of magnitude smaller than for water molecules.

\begin{table}
\begin{centering}
\begin{tabular}{|>{\centering}p{2.0cm}|>{\centering}p{2.0cm}|>{\centering}p{2.0cm}|>{\centering}p{2.0cm}|}
\hline 
& $N_{s}=96$ without embedding & $N_{s}=96$ with embedding & $N_{s}=256$ without embedding\tabularnewline
\hline 
\hline 
Average over embedded atoms  & 1.92 & 0.20 & 1.54\tabularnewline
\hline 
Average over other atoms  & 2.23 & 2.22 & 1.47\tabularnewline
\hline 
\end{tabular}
\par\end{centering}
\caption{\label{tab:forces_table}Average standard deviations of the atomic
force magnitudes, in ${\rm eV/\mathring{A}}$.}
\end{table}

The results show that embedding significantly reduces the stochastic
error of the accelerations for the selected (i.e., embedded) atoms.
As expected, when the same number of stochastic orbitals is used,
the standard deviation averaged over the non-embedded atoms remains
the same. Meanwhile, due to the good description of the embedded atoms
by the deterministic atomic basis, the standard deviations of the
forces for those atoms decrease by one order of magnitude relative
to the no-embedding case. To achieve without embedding the same level
of accuracy in the forces , we will need 10,000 stochastic orbitals,
which would be very time consuming. 

\section{Summary and prospects}

In summary, we presented a stochastic embedding DFT method (se-DFT)
that significantly reduces the statistical errors in the forces for
the selected subgroup of atoms (i.e., the embedded atoms). Combined
with the favorable linear scaling of stochastic DFT, the method can
be applied to large systems of practical interests. Of course, as
it stands the method is not efficient for overall MD of the full system,
due to the large stochastic errors on the environment (i.e., non-embedded)
atoms; rather it is suitable for applications where information on
a selected region is desired.

The embedding approach presented here is very general and can be extended
in several directions. First, here we used an embedded space made
from low-level atomic basis functions; we can replace it by a more
general higher level basis. Further, we could economize and choose
in the $P$ basis only occupied eigenfunctions of the embedded part
in, e.g., a dielectric medium. There will be occasions where the best
basis would be energy selective, i.e., a few energy-selective molecular
eigenfunctions or a few energy selective eigenfunctions from a large
cluster would be best used.

A second direction is a combination of embedding with our previous
overlapping fragment technique (sf-DFT). That method reduces the statistical
error of stochastic DFT calculations\cite{neuhauser2014communication,arnon2017equilibrium}
for all atoms, typically by up to an order of magnitude; specifically,
instead of stochastically sampling the full density, we sample the
difference between the full density and a zeroth-order density, $\rho(\boldsymbol{r})$
which is a solution of a simple zeroth-order Hamiltonian $H_{0}$
(e.g., that of overlapping fragments). Specifically:

\begin{equation}
\rho(\boldsymbol{r})=\rho_{0}(\boldsymbol{r})+\left\{ |\langle\boldsymbol{r}|\Theta^{\frac{1}{2}}|\xi\rangle|^{2}-|\langle\boldsymbol{r}|\Theta_{0}^{\frac{1}{2}}|\xi\rangle|^{2}\right\} _{\xi},\label{eq:Ovelapping_fragment}
\end{equation}
where $\Theta_{0}\equiv\Theta(\mu_{0}-H_{0})$ and $\mu_{0}$ is arbitrary.
It is clear that this overlapping-fragment definition can be further
extended by inserting projection operators as done earlier in the
paper for the original stochastic DFT method. In a future paper we
will examine whether a combination of overlapping fragments with embedding
(i.e., combining se-DFT with sf-DFT) reduces the errors in the forces
of the embedded part even further than either method alone. 

A third direction is for methods other than DFT. For example embedding
is applicable for the sub-linear scaling stochastic TDDFT method developed
by us \cite{gao2015sublinear}. It is straightforward to see that
the main embedding equation, Eq. (\ref{eq:rho_mu_embedded}), follows
straightforwardly to TDDFT except that now all quantities (the density
and the deterministic and stochastic orbitals) are time dependent.
Such a time-dependent method has the desired property that the same
Hamiltonian guides both the deterministic and stochastic function.
An embedding-TDDFT method would be applicable to study the change
in optical properties of chromophores due the the presence of solvent
molecules, where we can use only a few stochastic orbitals to sample
the solvent molecules, while the chromophore will be treated with
embedding. This direction would be explored in a future paper.

\section*{Acknowledgments}

D.N. acknowledges support from the NSF, Grant CHE-1763176. E.R. acknowledges
support from the Department of Energy, Photonics at Thermodynamic
Limits Energy Frontier Research Center, under grant number DE-SC0019140.
R.B. acknowledges support from the US-Israel Binational Science foundation
under the BSF-NSF program, Grant No. 2015687. 

The work also used resources of the National Energy Research Scientific
Computing Center (NERSC), a U.S. Department of Energy Office of Science
User Facility operated under Contract No. DE-AC02-05CH11231.  The
calculations were performed as part of the XSEDE\cite{towns2014xsede}
computational Project No. TG-CHE170058.

\section*{Appendix}

Here we give a simple demonstration of why embedding should improve
the statistics. Say that the overall problem has two spaces that are
essentially separate, so each eigenstate $\phi_{i}$ belongs to either
the $A$ or $B$ subspaces. we choose the $P$ subspace spanned by
$\{\chi_{i}:i\in A\}$ such that it is close to the subspace spanned
by states in A: 
\begin{center}
\begin{equation}
P|\phi_{i}\rangle\simeq c_{i}^{o}|\phi_{i}\rangle,
\end{equation}
\par\end{center}

\noindent where 
\begin{center}
\begin{equation}
c_{i}^{o}\approx\left\{ \begin{array}{cl}
1 & i\in A\\
0 & i\in B
\end{array}.\right.
\end{equation}
\par\end{center}

The projected filtered stochastic orbital $\bar{\xi}_{\mu}(\boldsymbol{r})$
is then given by a linear combination: 
\begin{center}
\begin{equation}
\bar{\xi}_{\mu}(\boldsymbol{r})=\sum_{i\le N_{{\rm Occ}}}(c_{i}-c_{i}^{o})\phi_{i}(\boldsymbol{r}).
\end{equation}
\par\end{center}

For states belonging to A, instead of sampling $c_{i}$ we are sampling
$(c_{i}-c_{i}^{o})$ with average $(1-c_{i}^{o})\approx0$ and a much
smaller standard deviation.

\bibliographystyle{unsrt}
\bibliography{ref}

\end{document}